\def\year{2018}\relax
\documentclass[letterpaper]{article} 
\usepackage{aaai18}  
\usepackage{times}  
\usepackage{helvet}  
\usepackage{courier}  
\usepackage{url}  
\usepackage{graphicx}  
\usepackage{amsmath}
\usepackage{amssymb}
\usepackage{booktabs}
\usepackage{multirow}
\usepackage{arydshln}

\DeclareMathOperator*{\argmax}{arg\!\max}
\usepackage{subfigure}
\usepackage{graphicx}

\DeclareMathOperator*{\argmin}{arg\,min}

\usepackage[english]{babel}

\makeatletter
  \newcommand\figcaption{\def\@captype{figure}\caption}
  \newcommand\tabcaption{\def\@captype{table}\caption}
\makeatother

\begin{document}
%

\title{Mesh-based Autoencoders for Localized Deformation Component Analysis}
\author{Qingyang Tan\textsuperscript{1,2}, Lin Gao\textsuperscript{1}\thanks{Corresponding Author}, Yu-Kun Lai \textsuperscript{3}, Jie Yang \textsuperscript{1,2} \and Shihong Xia\textsuperscript{1}\\
\textsuperscript{1}Beijing Key Laboratory of Mobile Computing and Pervasive Device, \\
Institute of Computing Technology, Chinese Academy of Sciences \\
\textsuperscript{2}School of Computer and Control Engineering, University of Chinese Academy of Sciences\\
\textsuperscript{3}School of Computer Science \& Informatics, Cardiff University\\
tanqingyang14@mails.ucas.ac.cn, \{gaolin, yangjie01, xsh\}@ict.ac.cn, LaiY4@cardiff.ac.uk
}

\maketitle

\begin{abstract}
Spatially localized deformation components are very useful for shape analysis and synthesis in 3D geometry processing. Several methods have recently been developed, with an aim to extract intuitive and interpretable deformation components. However, these techniques suffer from fundamental limitations especially for meshes with noise or large-scale deformations, and may not always be able to identify important deformation components.
In this paper we propose a novel mesh-based autoencoder architecture that is able to cope with meshes with irregular topology. We introduce sparse regularization in this framework, which along with convolutional operations, helps localize deformations.
Our framework is capable of extracting localized deformation components from mesh data sets with large-scale deformations and is robust to noise. It also provides a nonlinear approach to reconstruction of meshes using the extracted basis, which is more effective than the current linear combination approach. Extensive experiments show that our method outperforms state-of-the-art methods in both qualitative and quantitative evaluations.
\end{abstract}

\begin{figure*}
\begin{minipage}{0.42\textwidth}
\includegraphics[width = 1\textwidth]{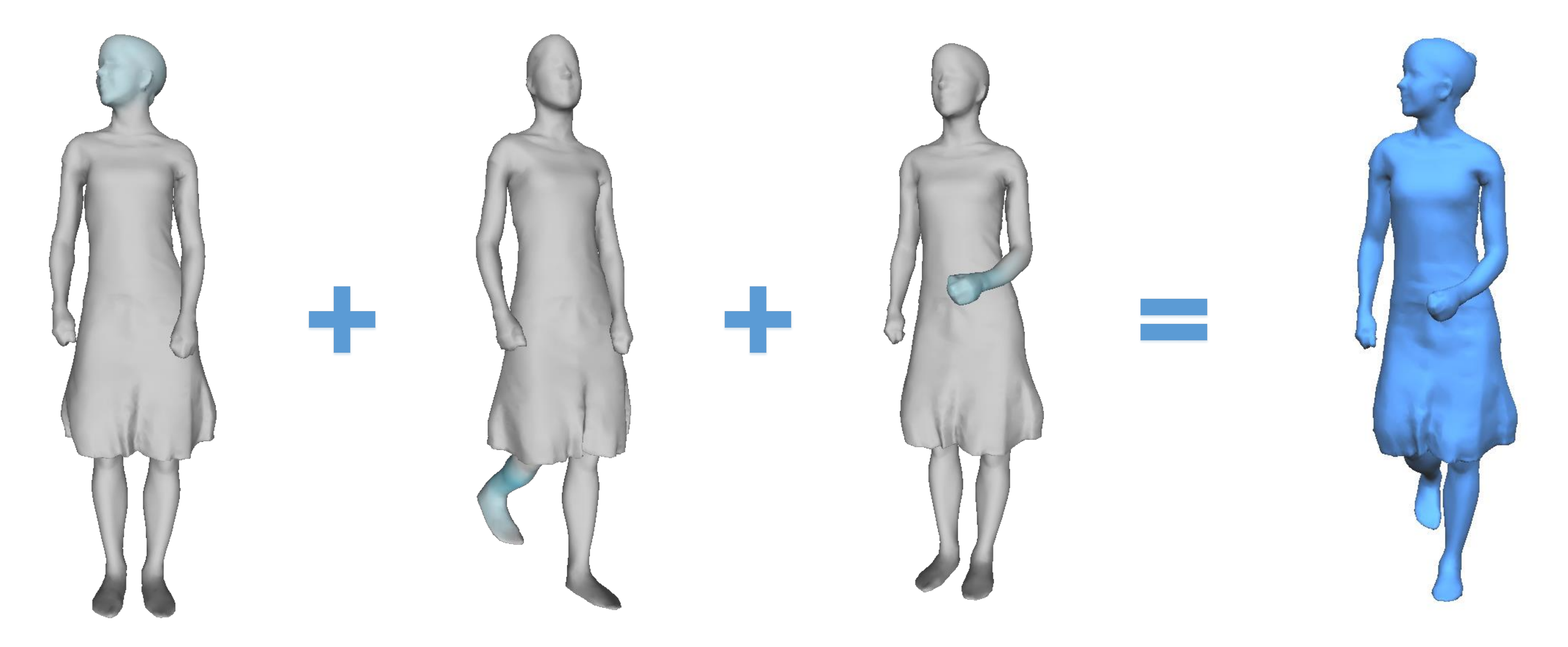}
\caption{Synthesized model by combining deformation components derived from the Swing dataset \shortcite{Vlasic2008} using our method with equal weights.}
\label{swingsynthesis}
\end{minipage}
\begin{minipage}{0.58\textwidth}
\includegraphics[width = 1\textwidth]{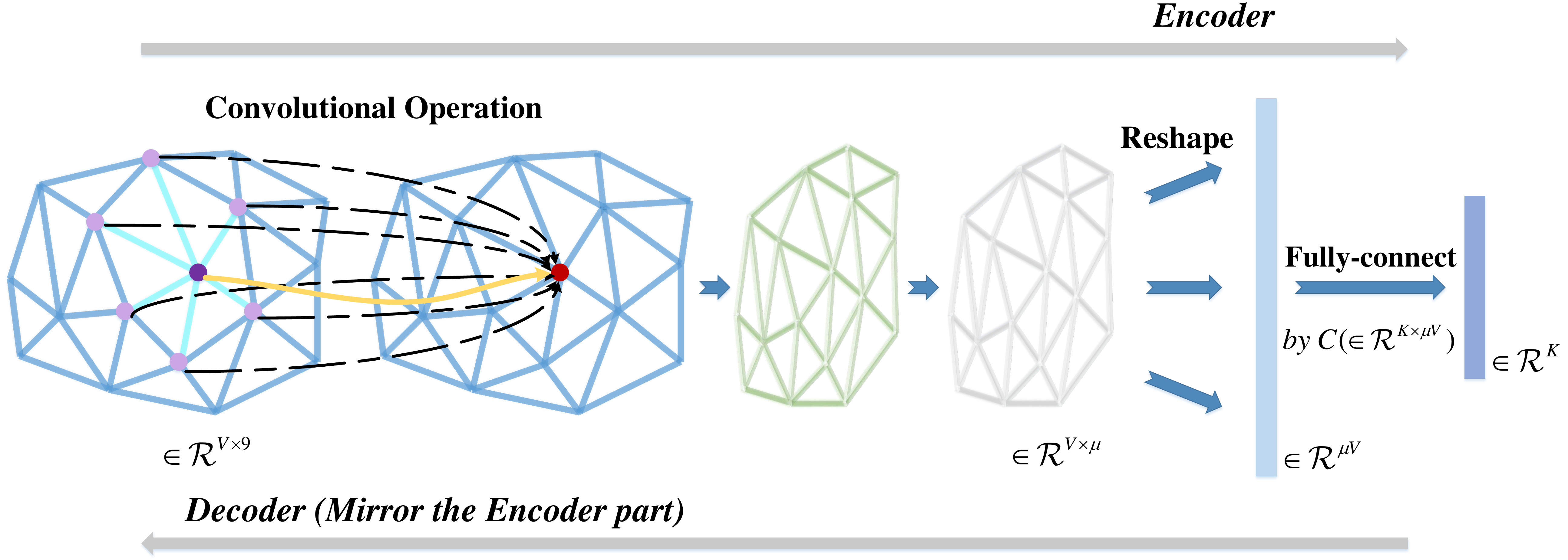}
\caption{The proposed network architecture.}
\label{pipeline}
\end{minipage}
\end{figure*}


\section{Introduction}
With the development of 3D scanning and modeling technology,  mesh data sets are becoming more and more popular. By analyzing these data sets with machine learning techniques, the latent knowledge can be exploited to advance geometry processing algorithms. In recent years, many research areas in geometry processing have benefited from this methodology, such as 3D shape deformation~\cite{Gao2016}, 
3D facial and human body reconstruction~\cite{Cao2015,Bogo2016}, shape segmentation~\cite{Guo2015},
 etc. For shape deformation and human reconstruction,  mesh sequences with different geometry and the same connectivity play a central role.
 Different geometric positions describe the appearance of the 3D mesh model while sharing the same vertex connectivity makes processing much more convenient. In such works, a key procedure is to build a low-dimensional control parametrization for the mesh data set, which provides a small set of intuitive parameters to control the generation of  new shapes. For articulated models such as human bodies, the rigging method embeds a skeleton structure in the mesh to provide such a  parametrization. However, the rigging operation is restrictive and does not generalize to other deformable shapes (e.g. faces). Parameterizing general mesh datasets which allows intuitive control in generating new shapes becomes an important and urgent research problem.

Early work extracted principal deformation components by using Principal Component Analysis (PCA) to reduce the dimensionality of the data set. However, such deformation components are global which do not lead to intuitive control. For example, when a user intends to deform the shape locally by specifying locally changed vertex positions as boundary conditions, the deformed shape tends to have unrelated areas deformed as well, due to the global nature of the basis.
To address this,  sparse localized deformation component (SPLOCS) extraction methods were recently  proposed~\cite{neumann2013sparse,huang,wang}. In these works the sparsity term is involved to localize deformation components within local support regions. However,
these previous works suffer from different limitations: as we will show later, \cite{neumann2013sparse,huang} cannot handle large-scale deformations,  and \cite{wang} is sensitive to noise which cannot extract the main deformation components robustly.
We propose a novel mesh-based autoencoder architecture to extract meaningful local deformation components. We represent deformations of shapes in the dataset based on a recent effective representation~\cite{Gao2017} which is able to cope with large deformations. We then build a CNN-based autoencoder to transform the deformation representation to encoding in a latent space. Each convolutional layer involves convolutional operations defined on the mesh with arbitrary topology in the form of applying the same local filter to each vertex and its 1-ring neighbors, similar to~\cite{duvenaud2015convolutional}. We then introduce sparsity regularization to the weights in the fully-connected layers to promote identifying sparse localized deformations.
The autoencoder structure ensures that the extracted deformation components are suitable for reconstructing high quality shape deformations.

Our main contributions are: 1) This is the first work that exploits CNN-based autoencoders for processing meshes with irregular connectivity. 2) Benefiting from sparse regularization and the nonlinear representation capability of autoencoders, our method is able to extract intuitive localized deformation components. It is able to deal with datasets with large-scale deformations, and is insensitive to noise. The method can extract important components even for challenging cases and generalizes well to reconstruction of unseen data. Extensive qualitative and quantitative experiments demonstrate that our method outperforms the state-of-the-art methods. We show an example of extracted deformation components (highlighted in blue) in Fig.~\ref{swingsynthesis}, which are then combined to synthesize a novel, plausible shape. The 
architecture of our proposed network is illustrated in Fig.~\ref{pipeline}.


\section{Related Work}\label{sec:related}

\paragraph{Principal Deformation Components Analysis.}
With the increasing availability of 3D shapes, analyzing shape collections is becoming more important.
Early work employs PCA to compress the mesh data set and extract global deformation components~\cite{Alexa2000}. The deformation components from the PCA are globally supported, which  is not intuitive for shape editing and deformation, especially when the user wants to deform the shape locally in the spatial domain~\cite{Havaldar2006}.
Sparse regularization is effective in localizing deformations~\cite{Gao2012}. 
However,  standard sparse PCA~\cite{zou2004} 
does not take spatial constraints into account and therefore the extracted deformation components do not aggregate in local spatial domains.
By incorporating spatial constraints, a sparsity term is employed to extract 
localized deformation components~\cite{neumann2013sparse,Bernard_cvpr}, 
which performs better than region-based PCA variants (clustered PCA)~\cite{Tena2011} in terms of extracting meaningful localized deformation components. However, it uses Euclidean coordinates which cannot represent shapes  with large rotations.
%
Later work addresses this limitation by using more advanced shape representations including deformation gradients~\cite{huang} and edge and dihedral angle representations~\cite{wang}. However, the former cannot cope with rotations larger than 180$^{\circ}$ which are very common in the animated mesh sequences, while the latter is not sensitive to the scale of the deformations which makes~\cite{wang} not robust to noise. Unlike existing methods, we propose to exploit mesh-based autoencoders with sparse regularization along with an effective deformation representation~\cite{Gao2017} to extract high-quality deformation components, outperforming existing methods.

\paragraph{Neural Network Applications for 3D Shapes.} Neural networks have achieved great success in different areas of computer science. Compared with 2D images, 3D shapes are more difficult to process, mainly due to their irregular connectivity and limited data availability. Nevertheless, some effort was made in recent years. For 3D object recognition, Su et al.~\shortcite{su2015multi} and Shi et al.~\shortcite{shi2015deeppano} represent 3D shapes using multi-view projections or converting them to panoramic views and utilize 2D CNNs. Maturana and Scherer~\shortcite{maturana2015voxnet} treat 3D shapes as voxels and extend 2D-CNNs to 3D-CNNs to recognize 3D objects.
In addition, Li et al.~\shortcite{li2015joint} analyze a joint embedding space of 2D images and 3D shapes.
Tulsiani et al.~\shortcite{tulsiani2016learning} abstract complex shapes using 3D volumetric primitives. For 3D shape synthesis, Wu et  al.~\shortcite{wu20153d} use deep belief networks to generate voxelized 3D shapes. Girdhar et al.~\shortcite{Girdhar16b} combine an encoder for 2D images and a decoder for 3D models to reconstruct 3D shapes from 2D input. Yan et al.~\shortcite{NIPS2016_6206} generate 3D models from 2D images by adding a projection layer from 3D to 2D.
Choy et al.~\shortcite{choy20163d} propose a novel recurrent network to map images of objects to 3D shapes. Sharma et al.~\shortcite{sharma2016vconv} train a volumetric autoencoder using noisy data with no labels for tasks such as denoising and completion. Wu et al.~\shortcite{3dgan} exploit the power of the generative adversarial network with a voxel CNN. In addition to  voxel representation, Sinha et al.~\shortcite{sinha2017surfnet} propose to combine ResNet and geometry images to synthesize 3D models. Li et al.~\shortcite{li_sig17} and Nash and Williams~\shortcite{nash2017shape} propose to use neural networks for encoding and synthesizing 3D shapes based on pre-segmented data.   
All the methods above for synthesizing 3D models are restricted by their representations or primitives adopted, which are not suitable for analyzing and generating 3D motion sequences with rich  details.

\paragraph{Convolutional Neural Networks (CNNs) on Arbitrary Graphs and Meshes.}
Traditional CNNs are defined on 2D images or 3D voxels with regular grids. Research has explored the potential to extend CNNs to irregular graphs by construction in the spectral domain~\cite{bruna2013spectral,
defferrard2016convolutional} or the spatial domain~\cite{niepert2016learning,duvenaud2015convolutional} focusing on spatial construction. Such representations are exploited in recent work~\cite{boscaini2016anisotropic,
Yi_2017_CVPR}
for finding correspondences or performing part-based segmentation on 3D shapes. Our method is based on spatial construction and utilizes this to build an autoencoder for analyzing deformation components.

\section{Feature Representation}\label{feature}

To represent large-scale deformations, we adapt a recently proposed deformation representation~\cite{Gao2017}. Given a dataset with $N$ shapes with the same topology, each shape is denoted as $S_{m}$, $m\in [1,\dots,N]$. {$\mathbf{p}_{m,i}\in \mathcal{R}^3$} is the $i^{\rm th}$ vertex on the $m^{\rm th}$ mesh model. The deformation gradient $\mathbf{T}_{m,i} \in \mathcal{R}^{3 \times 3}$  representing local shape deformations can be obtained by minimizing:
\begin{equation*}
\argmin_{\mathbf{T}_{m,i}} \sum_{j\in N(i)}c_{ij}\|(\mathbf{p}_{m,i}-\mathbf{p}_{m,j}) - \mathbf{T}_{m,i}(\mathbf{p}_{1,i}-\mathbf{p}_{1,j})\|_2^2.
\end{equation*}
where $c_{ij}$ is the cotangent weight and $N(i)$ is the index set of 1-ring neighbors of the $i^{\rm th}$ vertex.  By polar decomposition $\mathbf{T}_{m,i} = \mathbf{R}_{m,i}\mathbf{S}_{m,i}$, the affine matrix $\mathbf{T}_{m,i} \in \mathcal{R}^{3 \times 3}$ can be decomposed into an orthogonal matrix $\mathbf{R}_{m,i}$ describing rotations, and a real symmetry matrix $\mathbf{S}_{m,i}$ for scale and shear deformations.
 The rotation matrix $\mathbf{R}_{m,i}$ can be
 rewritten as rotating around an axis $\boldsymbol{\omega}_{m,i}$ by an angle $\theta_{m,i}$.
However, the mapping from the axis-angle representation to rigid rotation is surjective but not one to one:
The rotation angles and axes in the set $\Omega_{m,i}$ correspond to one rigid rotation:
\begin{equation*}
\Omega_{m,i} = \left\{ (\boldsymbol{\omega}_{m,i},\theta_{m,i} + t\cdot 2\pi), (-\boldsymbol{\omega}_{m,i},-\theta_{m,i} + t\cdot 2\pi)  \right\}
\end{equation*}
where $t$ is an arbitrary integer. 
To overcome this, ~\cite{Gao2017} proposes a novel representation to select the unique and consistent axis-angle representation by solving a global optimization to minimize the differences between adjacent rotation axes and angles.

For each vertex $i$ of shape $m$, we obtain feature $q_{m,i} = \{r_{m,i}, s_{m,i}\} \in \mathcal{R}^9$ by extracting from matrices $\mathbf{R}_{m,i}$ and $\mathbf{S}_{m,i}$. To fit the scale of output activation function $tanh$ (explained later), we need to scale the feature values. Denote by $r_{m,i}^j$ and $s_{m,i}^j$ the $j^{\rm th}$ dimension of $r_{m,i}$ and $s_{m,i}$ respectively.
Separately for each dimension $j$, we linearly scale $r_{m,i}^j$ and $s_{m,i}^j$  from $[r_{min}, r_{max}]$ and $[s_{min}, s_{max}]$ to $[-0.95, 0.95]$ to acquire preprocessed $\widetilde{r_{m,i}^j}$ and $\widetilde{s_{m,i}^j}$,
where $r_{min} = \min_{m,i,j}r_{m,i}^j$, and $r_{max},\ s_{min},\ s_{max}$ are defined similarly.  Then, we have $X_{m,i} = \{\widetilde{r_{m,i}}, \widetilde{s_{m,i}}\}$ as the deformation feature for vertex $i$ of shape $m$.

\section{Network Architecture}\label{sec:network}
In this section, we present our framework including convolutional operations on irregular meshes, overall network structure, sparsity constraints and reconstruction loss.

\subsection{Convolutional Operation}
Our convolutional operation is extended from~\cite{duvenaud2015convolutional} originally used for chemical molecules as a graph.  In our representation, a mesh with irregular connectivity is the domain, and data vectors are associated with each vertex. For a convolutional layer, it takes input data  $x\in\mathcal{R}^{V\times d}$, where $V$ is the number of vertices, and $d$ is the dimension of input data, and produces output data  $y\in \mathcal{R}^{V\times d'}$  where $d'$ is the dimension of the output data.
Denote by $x_i$ is the $i^{\rm th}$ row of $x$ corresponding to vertex $i$. Let its 1-ring neighbor vertices be $n_{ij}\ j\in{1,2,\dots,D_i}$, and $D_i$ is the degree of vertex $i$. The convolutional operation is computed as:
\begin{equation}
y_i = W_{point}x_i + W_{neighbour}\frac{\sum_{j=1}^{D_i}  x_{n_{ij}}}{D_i} + b,
\end{equation}
where $W_{point}, {W_{neighbour}} \in \mathcal{R}^{d'\times d}$ are weights for the convolutional operation, and $b \in \mathcal{R}^{d'}$ is the bias of the layer.

\subsection{Network Structure}
The overall network is built based on the convolutional operation and with an autoencoder structure. The input to the encoder part is preprocessed features which are shaped as $X \in \mathcal{R}^{V\times9}$, where $9$ is the dimension of the deformation representation. Then we stack several convolutional layers with $tanh$ as the output activation function. We tested alternative functions like $ReLU$, but they performed worse in the quantitative analysis.
The number of layers and the dimension of each layer are dependent on $V$ and model numbers for different datasets. If the encoder part has more than one convolutional layer, the last convolutional layer will directly use linear output without any non-linear activation function to avoid overfitting.
The output from the last convolutional layer is reshaped as a vector $f \in \mathcal{R}^{\mu V}$, where $\mu$ is the output dimension of the last convolutional layer. We use $C \in \mathcal{R}^{K\times \mu V}$ to map the feature to the latent space  $z \in \mathcal{R}^K$ where $K$ is the dimension of the latent space:
\begin{equation}
z = Cf.
\end{equation}
To reconstruct the shape representation from $z$, we use the decoder, which basically mirrors the encoder steps.
We first use the transpose of $C$ to transfer from the latent space back to the feature space:
\begin{equation}
\widehat{f} = C^Tz.
\end{equation}
For the decoder convolutional layers, we use the transposed weights of the corresponding layer in the encoder, with all layers using the $tanh$ output activation function. The output of the whole network is $\widehat{X}\in \mathcal{R}^{V\times9}$ which has the identical dimension as the input and can be scaled back to the deformation representation~\cite{Gao2017} and used for reconstructing the deformed shape.

The tied weight formulation of the autoencoder makes it more like PCA, and we assume that $F \in \mathcal{R}^{N\times\mu V}$ is  assembled by stacking all the features $f$ extracted from the last convolutional layer for $N$ models in the dataset. Then, $C$ can be seen as $K$ deformation components of $F$, and $Z \in \mathcal{R}^{N\times K}$ stacks the latent representations of the $N$ models in the dataset, which is treated as combinational weights to reconstruct the shape.
\begin{figure*}[tb]
\centering
\begin{minipage}[t]{0.27\textwidth}
\centering
\subfigure[]{\includegraphics[width = \textwidth]{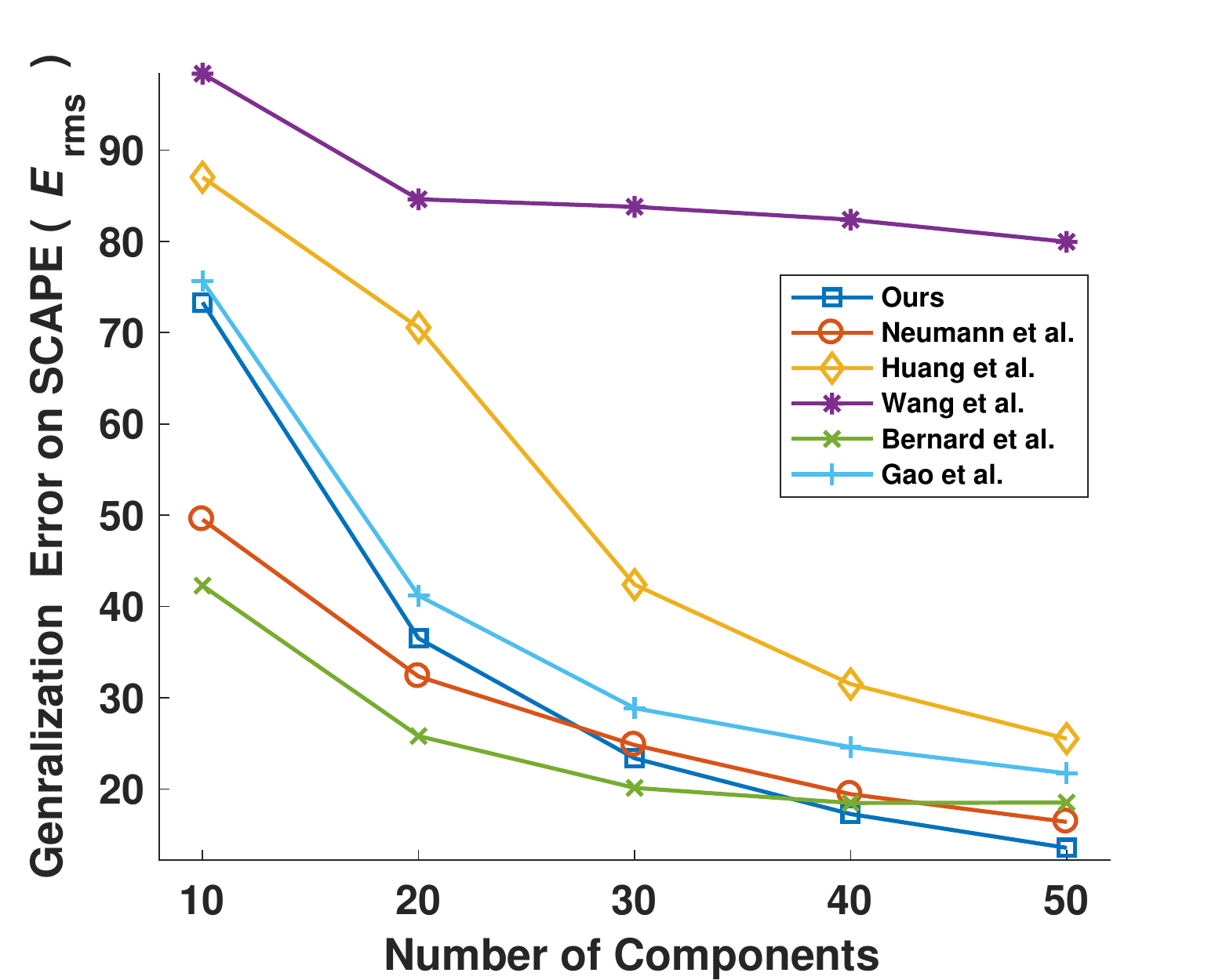}\label{scape_erms}}
\end{minipage}
\begin{minipage}[t]{0.27\textwidth}
\centering
\subfigure[]{\includegraphics[width = \textwidth]{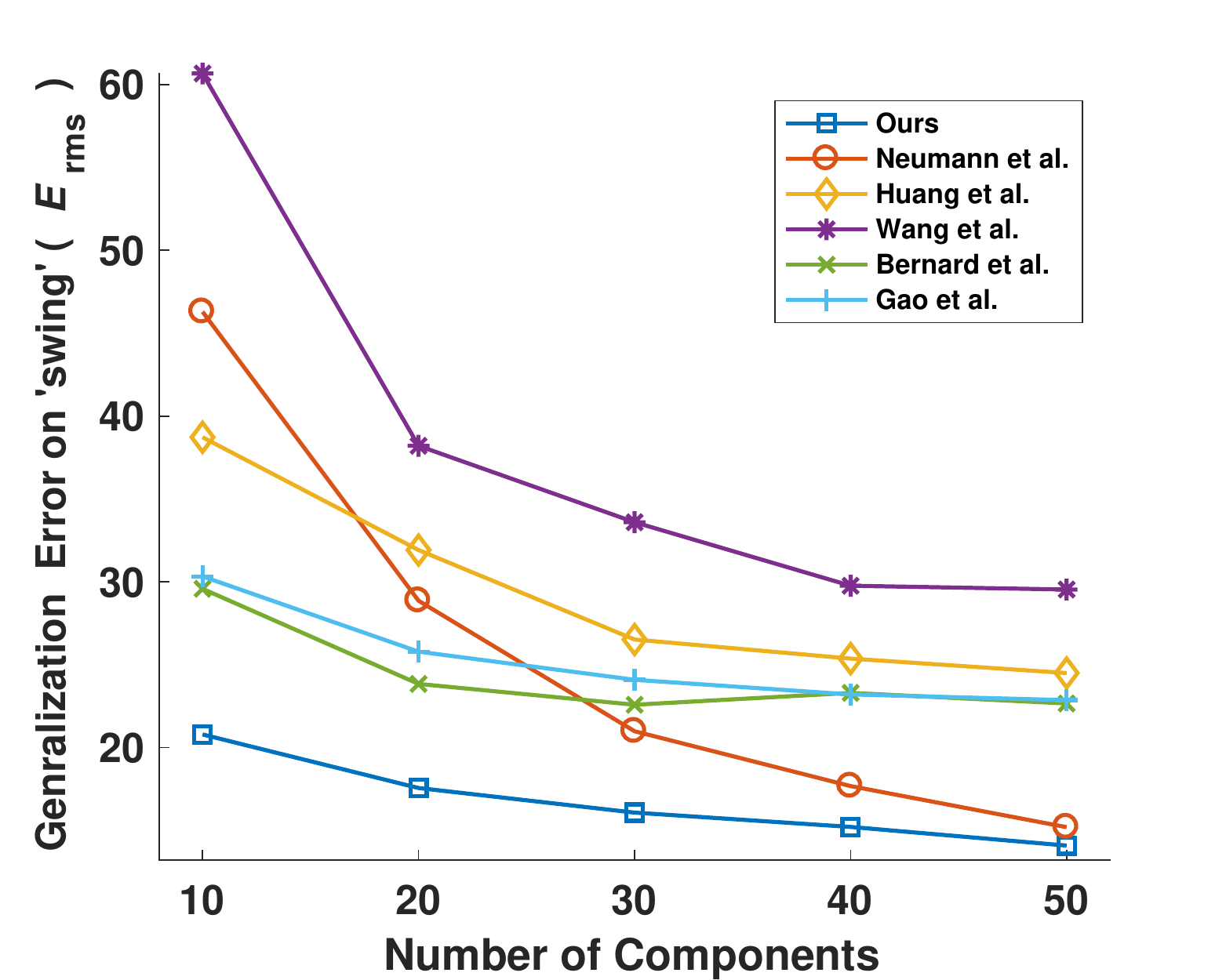}\label{swing_erms}}
\end{minipage}
\begin{minipage}[t]{0.27\textwidth}
\centering
\subfigure[]{\includegraphics[width = \textwidth]{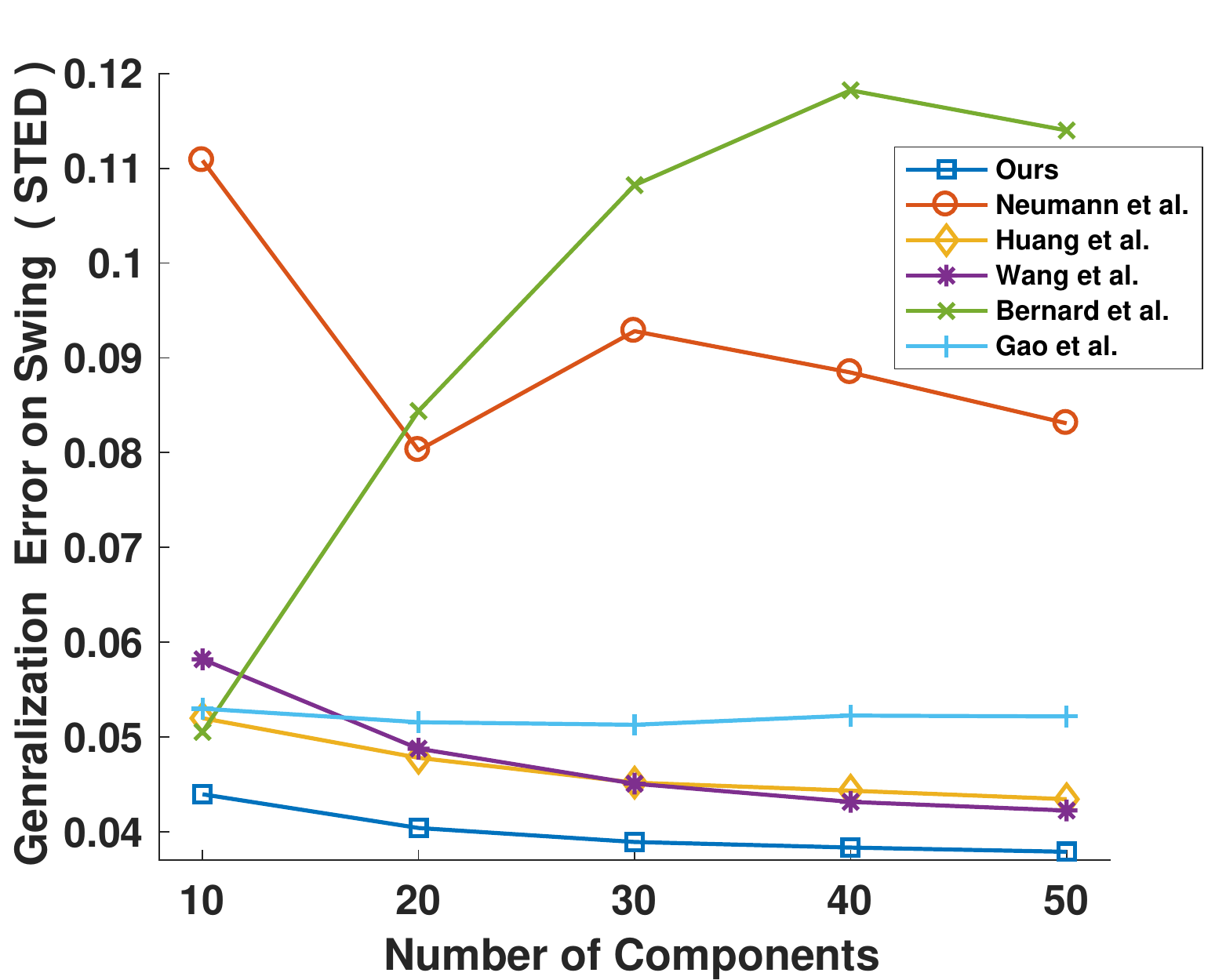}\label{swing_sted}}
\end{minipage}
\caption{Errors of applying our model to generate unseen data, using  (a) SCAPE \cite{SCAPE}, (b) (c) Swing  \cite{Vlasic2008} datasets. We use metrics $E_{rms}$ and STED with different component numbers. Our method outperforms other methods in all datasets and metrics even for limited training data.}
\label{gerror}
\end{figure*}

\begin{figure*}[t]
\begin{tabular}{lc}
\begin{minipage}{0.37\textwidth}
\centering
\includegraphics[width = 0.9\textwidth]{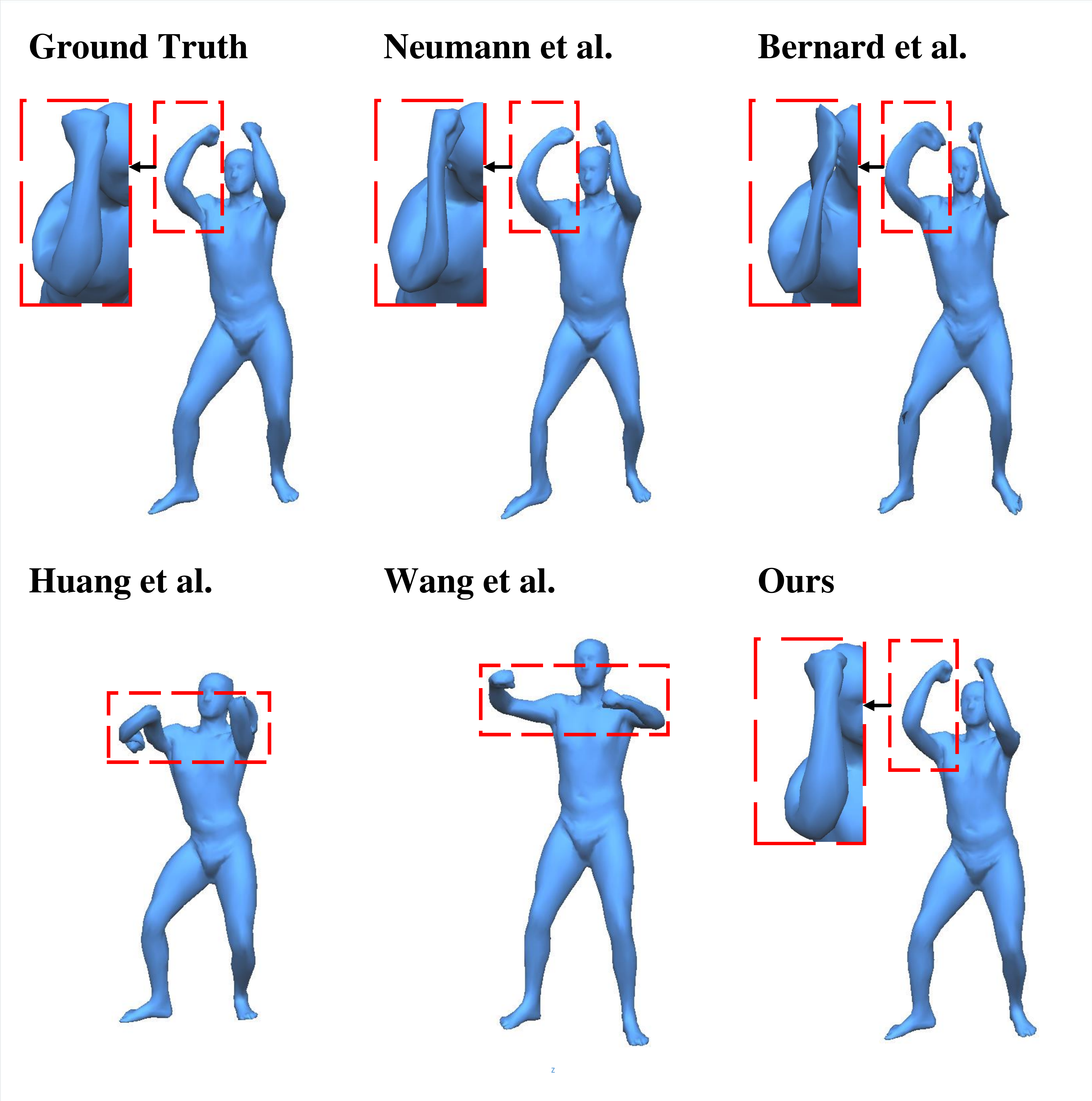}
\caption{Visual comparison of reconstruction results of the SCAPE dataset \shortcite{SCAPE}.}
\label{rscape}
\end{minipage}
\begin{minipage}{0.63\textwidth}
\centering
\small
\begin{tabular}{llccccc}
\toprule
\multirow{3}{*}{Dataset}&\multirow{3}{*}{Metric}&\multicolumn{5}{c}{Method} \\
\cmidrule(r){3-7}
 & &\multirow{2}{*}{Ours}  &Wang &Huang  & Neumann & Bernard\\
 & & &  et al. & et al. &et al. &et al.\\
\midrule
\multirow{2}{*}{Horse}& $E_{rms}$ & $12.9605$ & $29.6090$ & $18.0624$&$\mathbf{7.3682}$ & $20.1994$\\
\cmidrule(r){2-7}
 &$STED$& $\mathbf{0.04004}$ & $0.04332$ & $0.05273$ & $0.08074$ & $0.4111$\\
\midrule
\multirow{2}{*}{Face}& $E_{rms}$& $\mathbf{2.9083}$ & $8.5620$ & $12.3221$&$2.9106$ & $2.9853$ \\
\cmidrule(r){2-7}
 &$STED$& $\mathbf{0.007344}$ & $0.01320$ & $0.01827$ & $0.008611$ & $0.02662$ \\
\midrule
\multirow{2}{*}{Jumping}& $E_{rms}$& $\mathbf{24.4827}$ & $44.3362$ & $37.9915$&$29.3368$ & $49.9374$\\
\cmidrule(r){2-7}
 &$STED$& $\mathbf{0.04862}$ & $0.05400$ & $0.06305$ & $0.1268$ & $0.4308$\\
\midrule
\multirow{2}{*}{Humanoid} & $E_{rms}$& $\mathbf{3.4912}$ & $60.9925$ & $16.1995$ & $14.3610$ & $6.6320$\\
\cmidrule(r){2-7}
 &$STED$& $\mathbf{0.01313}$ & $0.03757$ & $0.02247$ & $0.07319$ & $0.04612$\\
\bottomrule
\end{tabular}
\tabcaption{Errors of applying our method to generate unseen data from Horse \cite{Sumner:2004:DTT:1015706.1015736}, Face \cite{zhang-siggraph2004-stfaces}, Jumping~\cite{Vlasic2008} and Humanoid datasets. We train all these methods with $50$ components.}
\label{moreevaluation}
\end{minipage}
\end{tabular}
\end{figure*}

\subsection{Sparsity Constraints and Reconstruction Loss}
Following the idea from \cite{neumann2013sparse}, we use group sparsity ($\ell_{2,1}$ norm) to urge deformation components to only capture local deformations. The constraints are added on $C$ as:
\begin{equation}
\Omega(C) = \frac{1}{K}\sum_{k=1}^K\sum_{i=1}^V\Lambda_{ik}\|C_k^{i}\|_2,
\end{equation}
where $C_k^{i}$ is the $\mu$-dimensional vector associated with  component $k$ of vertex $i$, and $\Lambda_{ik}$ is sparsity regularization parameters based on normalized geodesic distances:
\begin{equation}
\Lambda_{ik}=
\left\{
\begin{array}{ll}
0 & d_{ik}<d_{min} \\
1& d_{ik}>d_{max}\\
\frac{d_{ik} - d_{min}}{d_{max}-d_{min}}& {\rm otherwise}.
\end{array}
\right.
\end{equation}
\begin{figure}[tb]
\centering
\includegraphics[width = .45\textwidth]{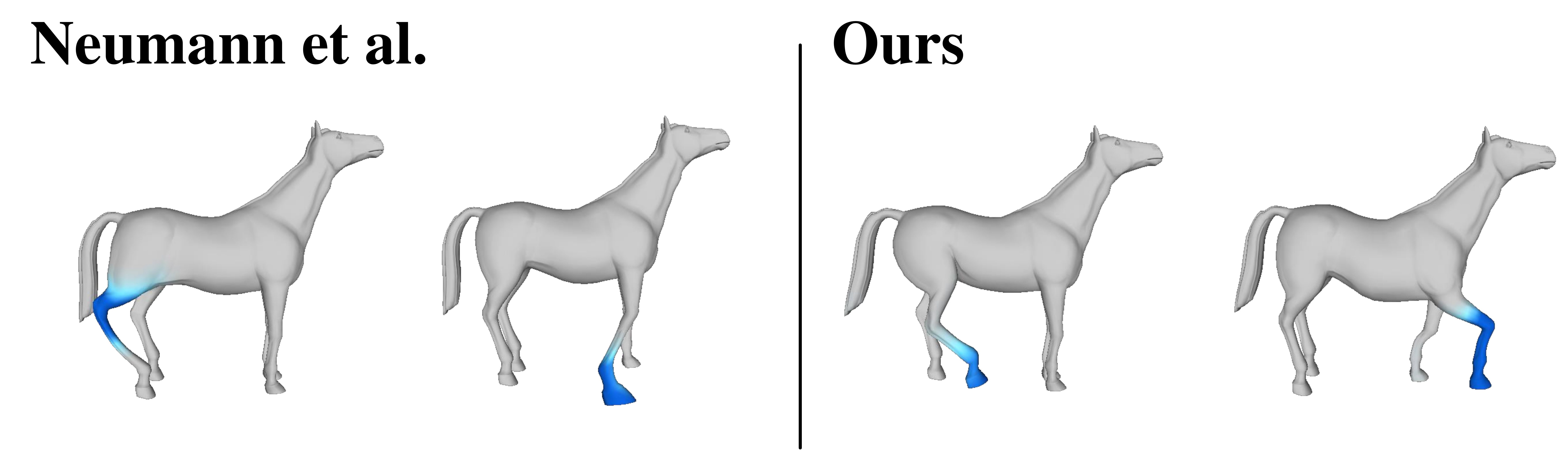}
\caption{Components of horse dataset \shortcite{Sumner:2004:DTT:1015706.1015736} extracted by \cite{neumann2013sparse} and our method.}
\label{horse}
\end{figure}
$d_{ik}$ denotes the normalized geodesic distance from vertex $i$ to the center point $c_k$ of component $k$ which is defined as:
\begin{equation}
c_k = \argmax_i\|C_k^i\|_2.
\end{equation}
$c_k$ will be updated after optimizing $C$ in each iteration. Intuitively, $\Lambda$ maps a geodesic distance to the range of $[0, 1]$ with distances out of the range of $[d_{min}, d_{max}]$ capped.
$d_{min}$ and $d_{max}$ are two tunable parameters, and control the size of deformation region of one component.
For most datasets, we use $d_{min} = 0.2$ and $d_{max}=0.4$.
To fit the training process of neural network, we precomputed all the geodesic distances between two vertices using \cite{Crane:2013:GH}, which are then normalized by the largest pairwise geodesic distance. 
\begin{figure*}[tb]
\centering
\begin{minipage}[t]{0.28\textwidth}
\centering
\subfigure[]{\includegraphics[width = 1\textwidth]{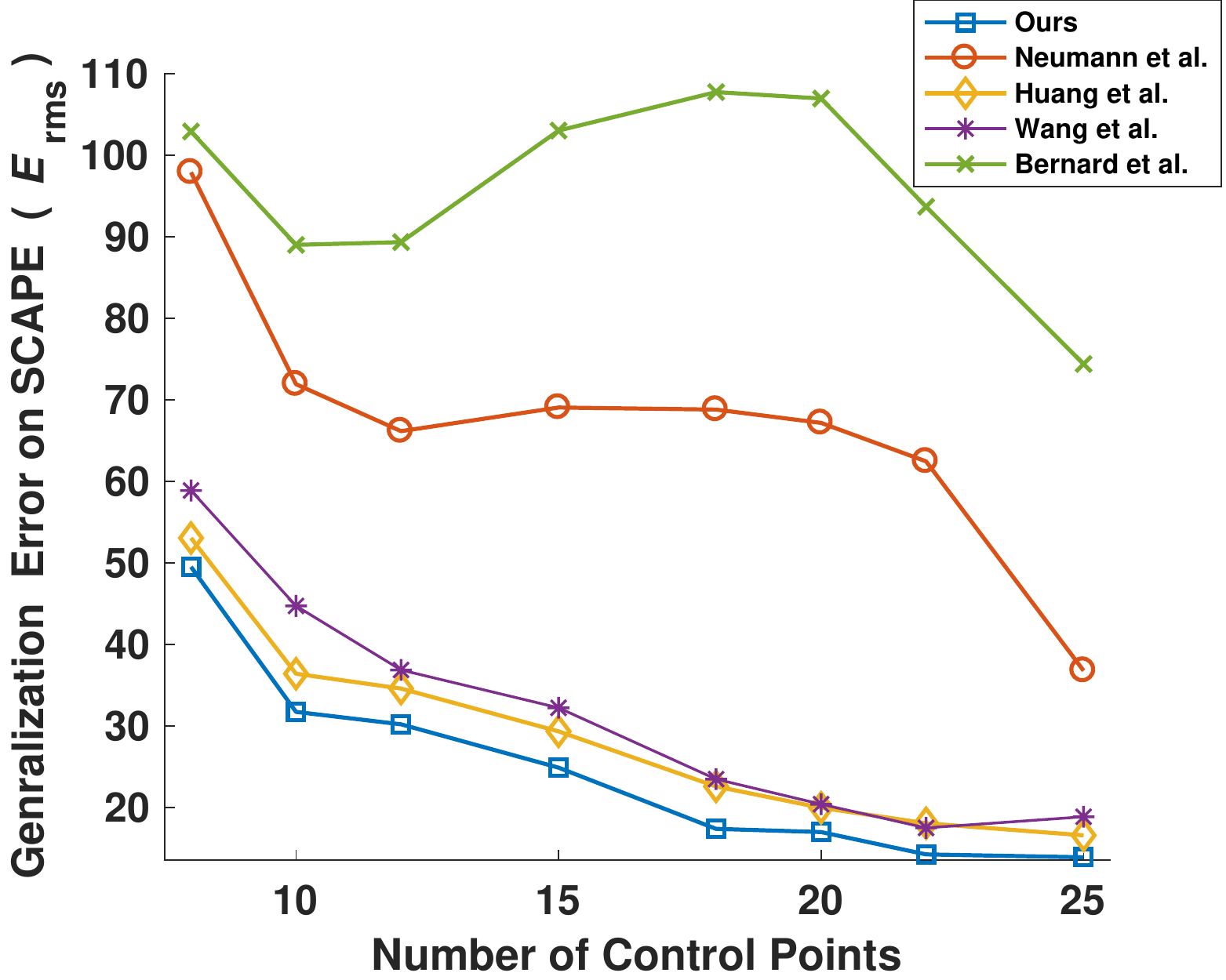}\label{sparse1}}
\end{minipage}
\begin{minipage}[t]{0.28\textwidth}
\centering
\subfigure[]{\includegraphics[width = 1\textwidth]{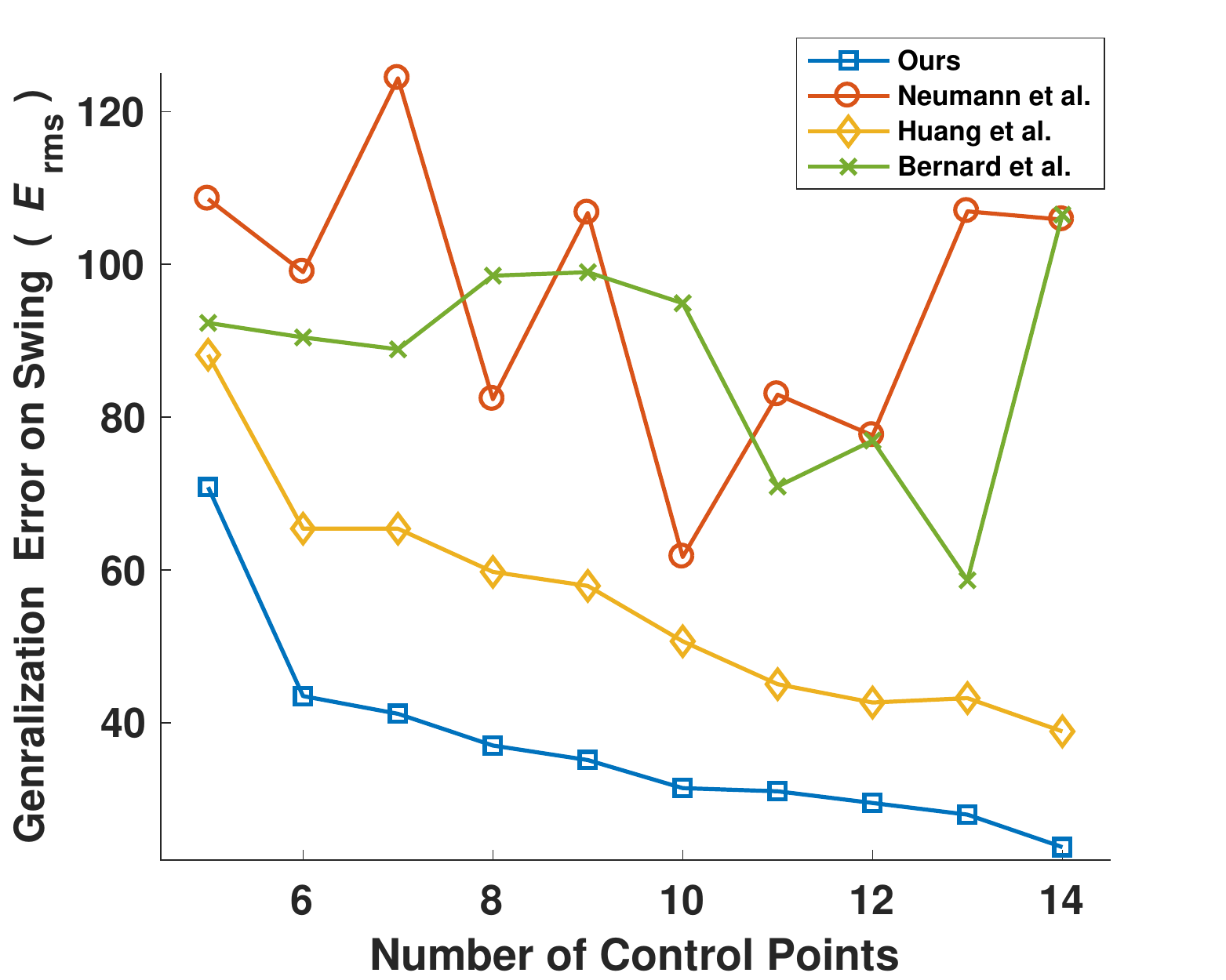}\label{sparse2}}
\end{minipage}
\begin{minipage}[t]{0.28\textwidth}
\centering
\subfigure[]{\includegraphics[width = 1\textwidth]{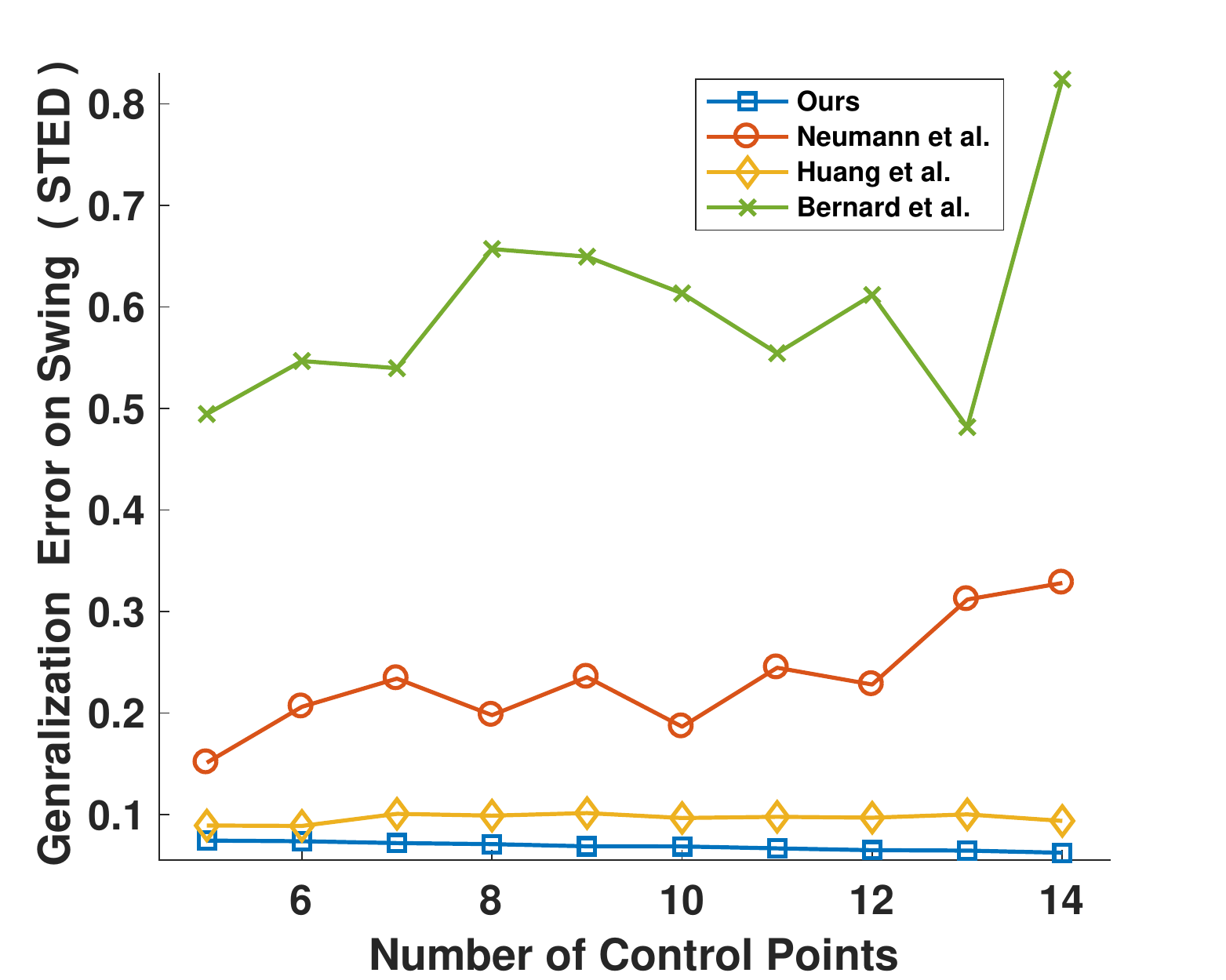}\label{sparse3}}
\end{minipage}
\caption{We use limited control points to reconstruct unseen data in the SCAPE \cite{SCAPE} and Swing~\cite{Vlasic2008} datasets, and report the generalization errors.}
\label{sparse_control}

\end{figure*}

\begin{figure*}[tb]
\centering
\includegraphics[width = .88\textwidth]{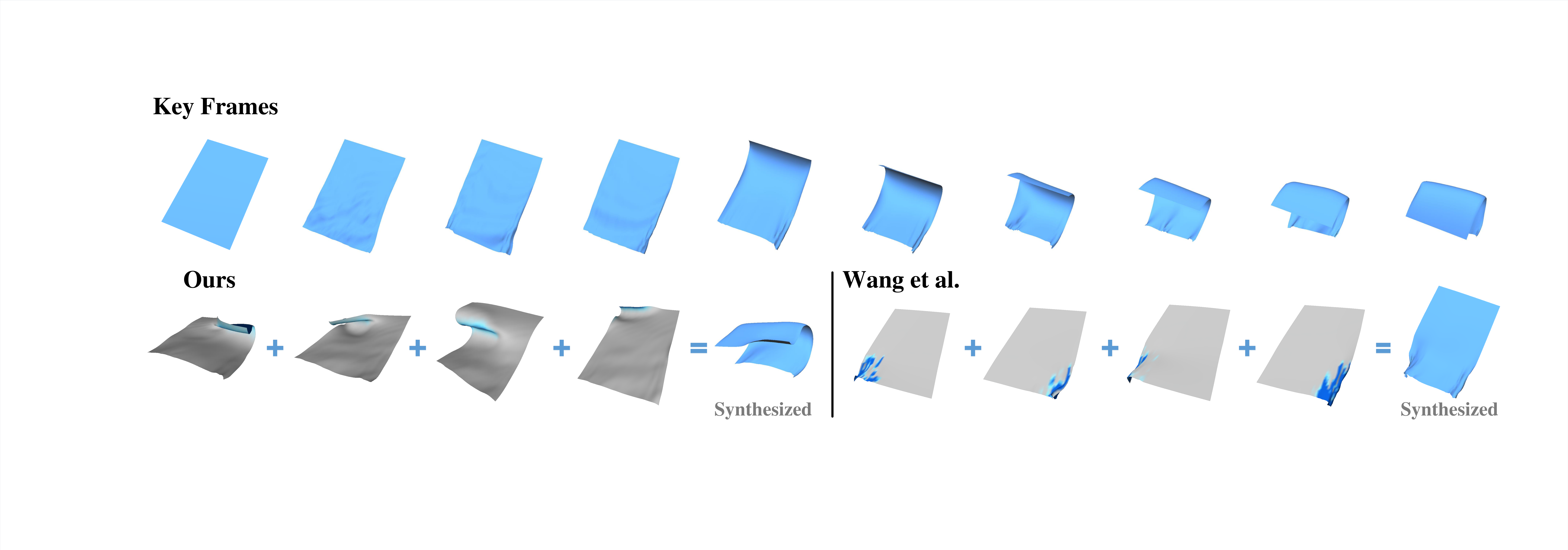}
\caption{Top row: key frames of a flag dataset we created through physical simulation. Bottom row: and the first four deformation components extracted by our method and \cite{wang}. We also present the synthesis results by combining the four components with equal weights, which shows our result is more plausible.}
\label{flag}
\end{figure*}

Since $C^TZ = (\frac{C^T}{\alpha})(\alpha Z) \quad \forall \alpha \neq 0$, to avoid trivial solutions with arbitrarily small $C$ values and arbitrary large $Z$ values, we also add constraints to $Z$ as a regularization term:
\begin{equation}
\mathcal{V}(Z) = \frac{1}{K}\sum_{j=1}^K (\max_m |Z_{jm}| - \theta),
\end{equation}
where $Z_{jm}$ is the $j^{\rm th}$ dimension of model $m$'s weight, and $\theta$ is a small positive number. We set $\theta = 5$ in all the experiments. We use Mean Square Error (MSE) to urge the network to reconstruct the representation of models, and the total loss function is:
\begin{equation}
\mathcal{L} = \frac{1}{N}\sum_{m=1}^N\|{\widehat{X}}_m - X_m\|_2^2 + \lambda_1\Omega(C)+ \lambda_2\mathcal{V}(Z),
\end{equation}
where ${\widehat{X}}_m$ and $X_m$ are input and output of model $m$ (data term), $\Omega(C)$ is the sparse  localized regularization. We set $\lambda_1 = \lambda_2 = 0.5$ in all the experiments. The whole network pipeline is illustrated in Fig. \ref{pipeline}. We use ADAM algorithm \cite{Kingma2015} and set the learning rate to be $0.001$ to train the network.


\section{Applications}\label{sec:application}
Once trained, the network can be used to perform many useful tasks, including dimensionality reduction, reconstruction, component analysis and shape synthesis. The first two applications are straightforward, so we now give details for performing the last two applications.

\subsection{Component Analysis}
The matrix $C$ corresponds to the localized deformation components. We assume the $r^{\rm th}$ model is the reference model (which can be the first model in the dataset) which has a latent vector $Z_r$.  To analyze the $i^{\rm th}$ deformation component, we calculate the minimum and maximum values of the $i^{\rm th}$ dimension of the embedding, denoted by $Z_{i_{min}} = \min_m Z_{i,m}$  and $Z_{i_{max}} = \max_m Z_{i,m}$. We can then obtain latent vectors $\widehat{Z}_{i_{min}}$ and $\widehat{Z}_{i_{max}}$ corresponding to the two extreme values of the $i^{\rm th}$ component by replacing the $i^{\rm th}$ component of $Z_r$ with  $Z_{i_{min}}$ and $Z_{i_{max}}$, respectively. Applying the vectors to the decoder produces the output mesh features $\widehat{X_{min}}$ and $\widehat{X_{max}}$. We work out the differences $\|\widehat{X_{min}} - X_r\|$ and $\|\widehat{X_{max}} - X_r\|$  and the one that has larger distance from the reference model $X_r$ is chosen as the representative shape for the $i^{\rm th}$ deformation component, with the corresponding latent vector denoted as $Z_{i_h}$. The displacement of each vertex feature indicates the strength of the deformation, which can be visualized to highlight changed positions.


\subsection{Shape Synthesis} 
To synthesize new models, the user can specify a synthesis weight $w_{s_i}$ for the $i^{\rm th}$ deformation component, and the deformed shape in the latent space can be obtained as:
\begin{equation}
{z_s}_i = {Z_i}_r + ({Z_i}_h - {Z_i}_r) \times {w_s}_i,
\end{equation}
where ${z_s}_i$ represents the $i^{\rm th}$ dimension of obtained weight $z_s$ in the latent space. Then, by feeding $z_s$ in as input to the decoder, the synthesized model feature can be obtained which can be used for reconstructing the synthesized shape.



\section{Experimental Results}\label{results}

\subsection{Quantitative Evaluation}
We compare the generalization ability of our method with several state-of-the-art methods, including original SPLOCS~\cite{neumann2013sparse}, SPLOCS with deformation gradients~\cite{huang}, SPLOCS with edge lengths and dihedral angles~\cite{wang}, SPLOCS with the feature from \cite{Gao2017} as used in this paper, and \cite{Bernard_cvpr}. We use SCAPE~\cite{SCAPE} and Swing~\cite{Vlasic2008} datasets to conduct main quantitative evaluation.




For the SCAPE dataset, we randomly choose $36$ models as the training set and the remaining $35$ models as the test set. After training, we compare the generalization error on the test set with different methods, using $E_{rms}$ (root mean square) error~\cite{CGF:CGF1602}. The results are shown in Fig. \ref{scape_erms}. Fig. \ref{rscape} shows the visual comparison of reconstruction results. For the Swing dataset, we randomly select one model from every ten models for training ($15$ models) and remaining for testing ($135$ models). We compare $E_{rms}$ error as well as $STED$ error~\cite{5416707} designed for motion sequences with a focus on `perceptual' error of models. The results are shown in Figs. \ref{swing_erms} and  \ref{swing_sted}. Note that since the vertex position representation cannot handle rotations well, the more components methods~\cite{neumann2013sparse,Bernard_cvpr} use, the more artifacts would be brought in the reconstructed models, thus $STED$ error may increase with more components. The results indicate that our method has better quantitative reconstruction results than other methods, with lower reconstruction errors when sufficient components are used. From the visual results, we can see that \cite{neumann2013sparse,Bernard_cvpr,huang} cannot handle large-scale rotations well and cannot reconstruct plausible models in such cases, while \cite{wang} can be affected by noise in the dataset and cannot recover some actions precisely. Our method does not have such drawbacks. Comparison with SPLOCS using~\cite{Gao2017} demonstrates that our autoencoder is effective, beyond the benefits from the representation.
We also experiment using linearly combined components derived from our method to reconstruct unseen models, and the errors of SPLOCS with~\cite{Gao2017}  are greater than our non-linear framework. For example, for the Swing dataset with 50 components, the $E_{rms}$ error is $25.2994$, and $STED$ error is $0.05214$ (whereas for our method these two errors are $14.0836$ and $0.03789$
). For the SCAPE dataset, the $E_{rms}$ error is $17.1754$, which is larger than our error $13.556$. This shows that our non-linear method can find intrinsic patterns and better fit the relationships between the latent space and feature domain. 

The major parameters in our method are $\lambda_1$ and $\lambda_2$, which are used for balancing regularization terms. For all the experiments, we set them to $0.5$. To verify the sensitivity of our method with these two parameters, we perform additional experiments to compare results when we change them in the range of $0.4$--$0.6$, and it does not greatly affect quantitative performance. An example is shown in Table \ref{diff_lambda}, which is performed on the SCAPE dataset with 50 components.

Following the previous experiment, $50$ components are generally sufficient to fit data well for all the methods, which are therefore used in the following comparative experiments. The results
are summarized in Table~\ref{moreevaluation}. All the datasets we use here can be seen as motion sequences, so we use the same training-test split used for the Swing dataset, and use the two metrics to evaluate errors. Although for the Horse dataset~\cite{Sumner2005}, the method~\cite{neumann2013sparse} has a lower $E_{rms}$ error than our method, their method cannot cope with such dataset with large deformations and suffers from artifacts.
The components extracted by our method and \cite{neumann2013sparse} are shown in Fig. \ref{horse}.

Meanwhile, to quantitatively compare the sparse control ability of these methods, we randomly select a few points on the mesh and test the ability of each method  to recover the whole mesh through these limited points. This situation is similar to  the scenario that users put limited control points on significant joints to acquire models with meaningful actions. To obtain control points evenly distributed on the mesh surface, we randomly choose the first point, and then use Voronoi sampling to acquire the other points. We test the results on SCAPE and Swing datasets. For both methods, we choose 50 components, and for \cite{neumann2013sparse} and \cite{Bernard_cvpr}, we solve the reconstruction problem directly using the limited points, while for the other methods, we use data-driven deformation with the extracted components. The results in  Fig.~\ref{sparse_control} show that our method performs well, consistently with smallest errors in both metrics. The datasets we use in this experiment contain a great amount of rotation. Therefore,  using limited control points may not allow the components extracted by \cite{neumann2013sparse} and \cite{Bernard_cvpr} to recover the whole mesh, resulting in large fluctuations in the error curves.

\begin{figure}[tb]
\centering
\includegraphics[width = .33\textwidth]{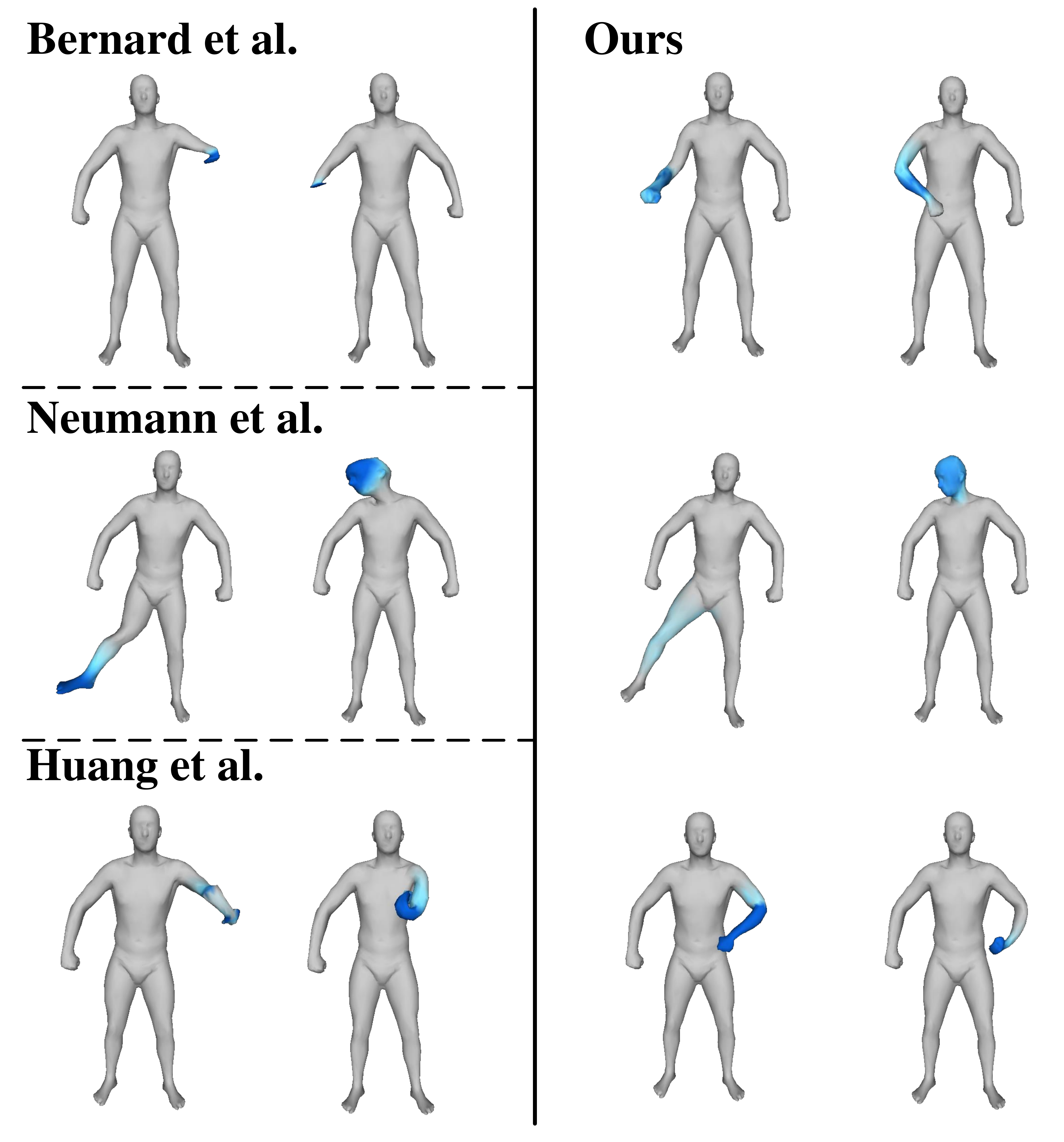}
\caption{Comparison of deformation components located in similar areas, which are extracted by different methods.}
\label{scape}
\end{figure}

\begin{figure}[tb]
\centering
\includegraphics[width = 0.48\textwidth]{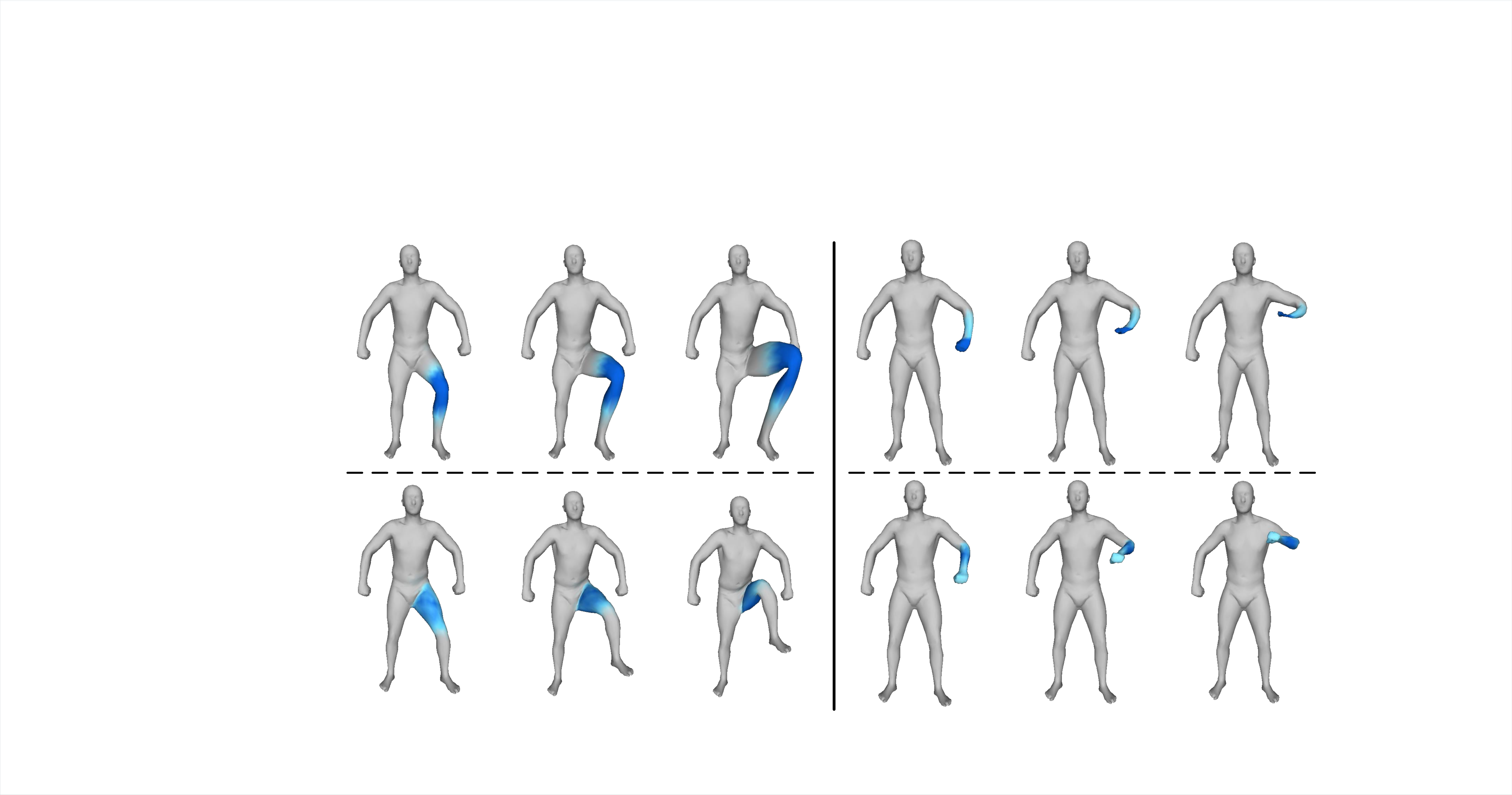}
\caption{Synthesis results with different components of the SCAPE dataset \shortcite{SCAPE}:
The left group contains the components about lifting the left leg extracted by \citeauthor{neumann2013sparse} (first row) and our method (second row) with weights $0.5$, $1.0$ and $1.5$. The right group contains the components about lifting the left arm extracted by \citeauthor{huang} (first row) and our method (second row) with weights $0.3$, $0.6$ and $0.9$.}
\label{scapeedit}
\end{figure}

\begin{figure}[tb]
\centering
\includegraphics[width = .32\textwidth]{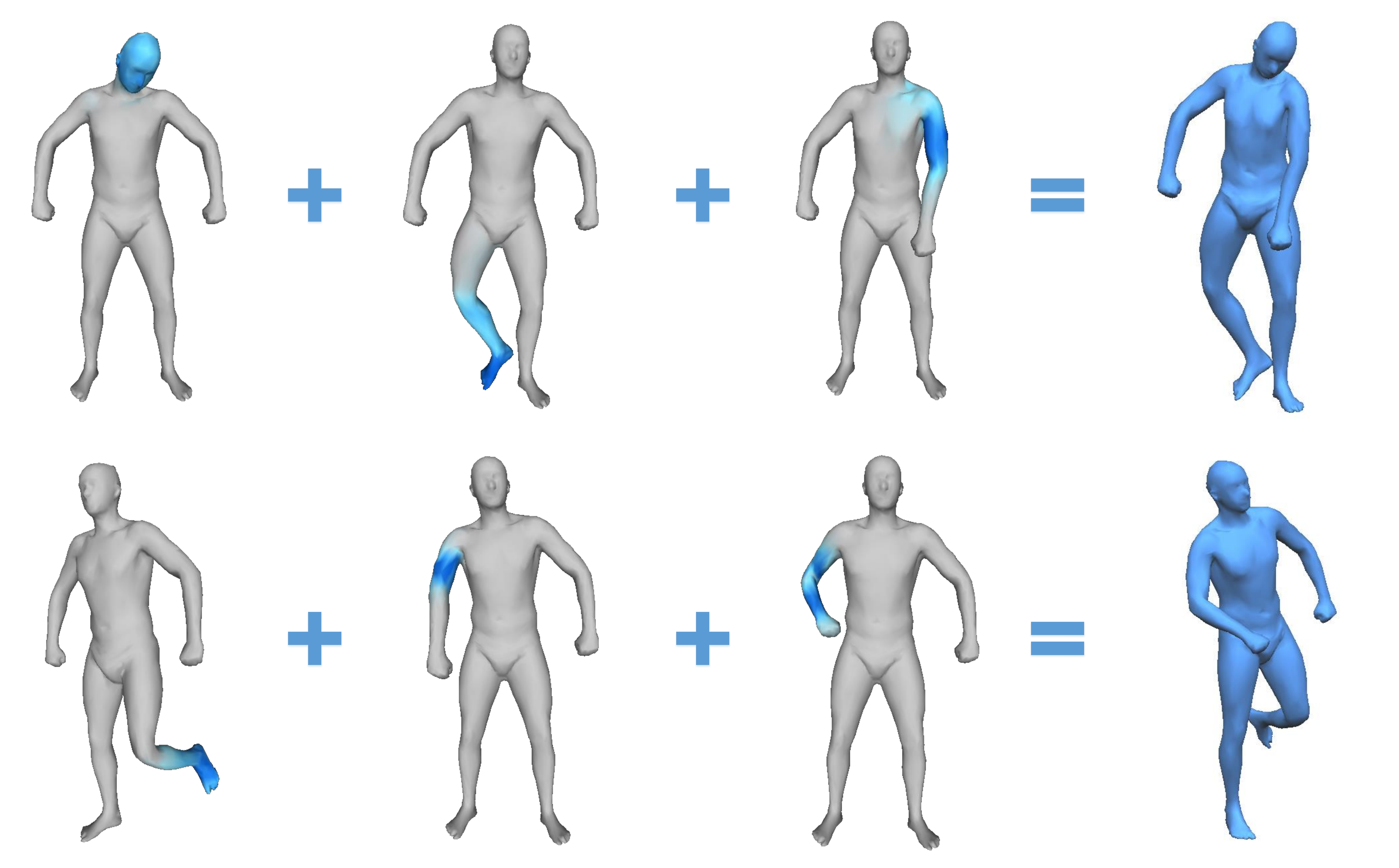}
\caption{Synthesized models based on components derived from SCAPE dataset \shortcite{SCAPE} by our method.}
\label{scapesynthesis}
\end{figure}

\begin{figure}[tb]
\centering
\includegraphics[width = .48\textwidth]{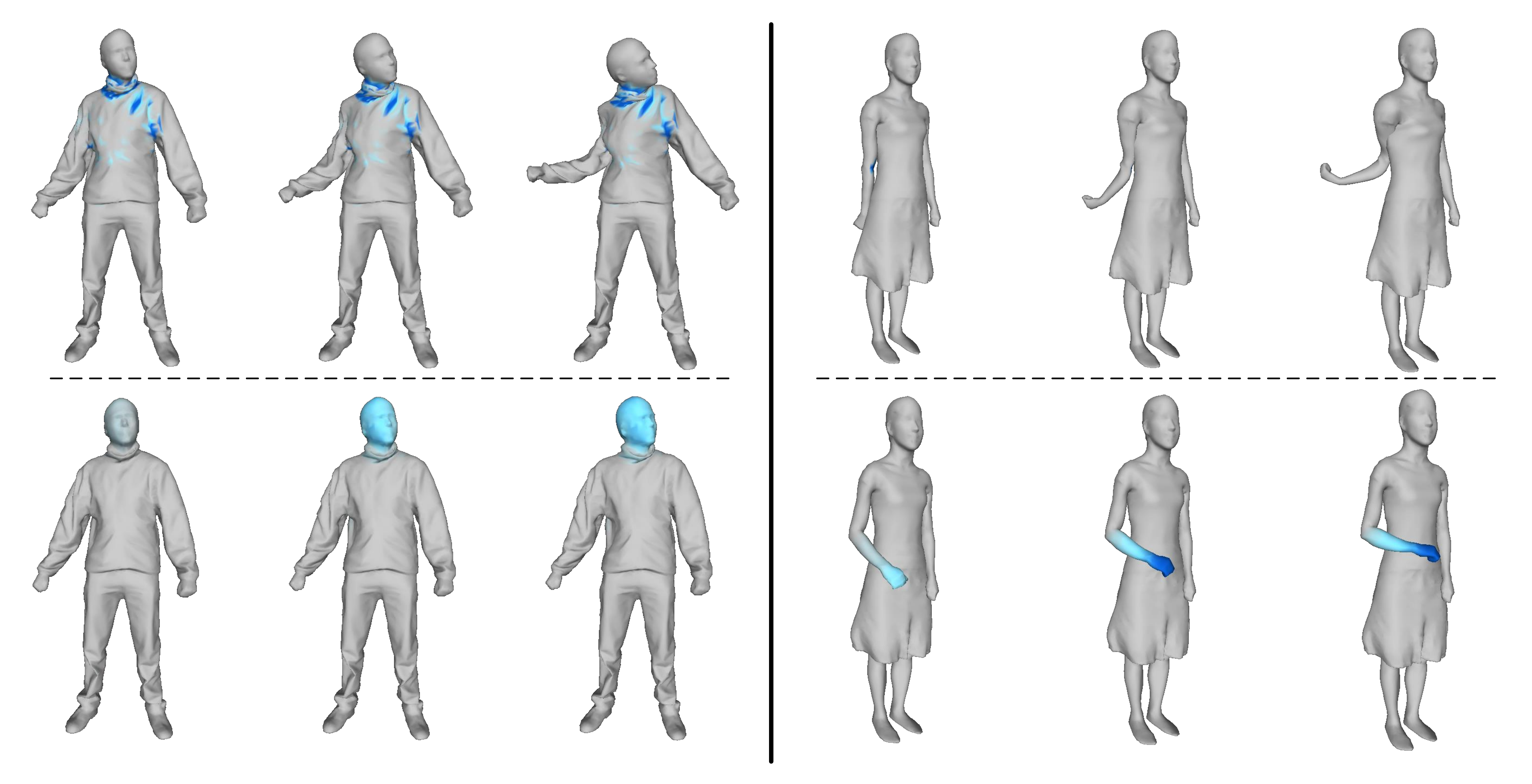}

\caption{Synthesis results with components of Jumping and Swing datasets~\shortcite{Vlasic2008}:
The left group contains the components about shaking head extracted by \citeauthor{wang} (first row) and our method (second row) with weights $0.4$, $0.8$ and $1.2$.
The right group contains the components about lifting the right arm extracted by \citeauthor{wang} (first row) and our method (second row) with weights $0.6$, $0.9$ and $1.2$.
}
\label{swingedit}
\end{figure}

\begin{table}[t]
\centering
\begin{tabular}{lcccc}
\toprule
$(\lambda_1, \lambda_2)$ & $(0.4, 0.4)$ & $(0.6, 0.6)$ & $(0.4, 0.6)$ & $(0.6, 0.4)$\\
\midrule
$E_{rms}$ & $14.1763$ & $13.6429$ & $14.2645$& $14.093$\\
\bottomrule
\end{tabular}
\caption{Reconstruction error for unseen data from SCAPE \cite{SCAPE} when different $\lambda_1$ and $\lambda_2$ are used. The default values are both $0.5$, and the error is $13.556$.}
\label{diff_lambda}
\end{table}

\subsection{Qualitative Evaluation}

\subsubsection{Flag Dataset.}

To verify our method's ability to capture primary deformation components even when there is significant noise, we test on a flag dataset created by physical simulation and compare  our method with \cite{wang}. For both methods, we extract $20$ components, and the first four components along with the key frames of the dataset are shown in Fig. \ref{flag}. Our method is able to extract the main movements (large-scale swinging of the flag), and separate local movements in the left and right parts of the flag. The synthesized result with the four components is reasonable. However, \cite{wang} only captures the noise around the corner of flags, and the reconstructed shape does not capture the true deformation.

\subsubsection{SCAPE Dataset.}
We compare our results on the SCAPE dataset with \cite{neumann2013sparse,huang,Bernard_cvpr}. The corresponding components extracted by our method and the other methods are shown in Fig. \ref{scape}, and two groups of components about lifting the left leg (extracted by our method and \citeauthor{neumann2013sparse}) and left arm (extracted by our method and \citeauthor{huang}) with different weights are shown in Fig. \ref{scapeedit}. These justify that our method can handle large-scale rotation better than the other methods without artifacts like irrational amplification and shrinkage. Our proposed method also has powerful synthesis ability.  We show synthesis results by combining several different deformation components in Fig. \ref{scapesynthesis}.

\subsubsection{Swing and Jumping Datasets.}
For Swing and Jumping datasets from \cite{Vlasic2008}, we 
align all the models and then train the network. The synthesis results of our method are compared with those of \cite{wang} in Fig. \ref{swingedit}. The first group of components are about shaking head to left from the Jumping dataset. Our method focuses on the movement of the head and can produce reasonable models, while models generated by \cite{wang} are disturbed by the clothes, and have artifacts of arm structure. The second group of models are about lifting the left arms from the Swing dataset, \cite{wang} even finds wrong direction for this movement. We show synthesis results by combining three different 
components in Fig. \ref{swingsynthesis}.

\section{Conclusion}\label{conclusions}
In this paper, we propose a novel CNN based autoencoder on meshes to extract localized deformation components. Extensive quantitative and qualitative evaluations show that our method is effective, outperforming state-of-the-art methods.

\section{Acknowledgments}
  This work was supported by the National Natural Science Foundation of China (No.61502453, No.61772499 and No.61611130215), Royal Society-Newton Mobility Grant (No. IE150731), the Science and Technology Service Network Initiative of Chinese Academy of Sciences (No. KFJ-STS-ZDTP-017)
  and the NVIDIA hardware donation.
\fontsize{9.0pt}{10.0pt}
\selectfont
\bibliography{sparse}

\begin{thebibliography}{}

\bibitem[\protect\citeauthoryear{Alexa and Muller}{2000}]{Alexa2000}
Alexa, M., and Muller, W.
\newblock 2000.
\newblock {Representing Animations by Principal Components}.
\newblock {\em Comp. Graph. Forum}.

\bibitem[\protect\citeauthoryear{Anguelov \bgroup et al\mbox.\egroup
  }{2005}]{SCAPE}
Anguelov, D.; Srinivasan, P.; Koller, D.; Thrun, S.; Rodgers, J.; and Davis, J.
\newblock 2005.
\newblock {SCAPE}: shape completion and animation of people.
\newblock {\em ACM Trans. Graph.} 24(3):408--416.

\bibitem[\protect\citeauthoryear{Bernard \bgroup et al\mbox.\egroup
  }{2016}]{Bernard_cvpr}
Bernard, F.; Gemmar, P.; Hertel, F.; Goncalves, J.; and Thunberg, J.
\newblock 2016.
\newblock Linear shape deformation models with local support using graph-based
  structured matrix factorisation.
\newblock In {\em CVPR},  5629--5638.

\bibitem[\protect\citeauthoryear{Bogo \bgroup et al\mbox.\egroup
  }{2016}]{Bogo2016}
Bogo, F.; Kanazawa, A.; Lassner, C.; Gehler, P.; Romero, J.; and Black, M.~J.
\newblock 2016.
\newblock Keep it {SMPL}: Automatic estimation of {3D} human pose and shape
  from a single image.
\newblock In {\em ECCV}.

\bibitem[\protect\citeauthoryear{Boscaini \bgroup et al\mbox.\egroup
  }{2016}]{boscaini2016anisotropic}
Boscaini, D.; Masci, J.; Rodol{\`a}, E.; Bronstein, M.~M.; and Cremers, D.
\newblock 2016.
\newblock Anisotropic diffusion descriptors.
\newblock In {\em Comp. Graph. Forum}, volume~35,  431--441.

\bibitem[\protect\citeauthoryear{Bruna \bgroup et al\mbox.\egroup
  }{2013}]{bruna2013spectral}
Bruna, J.; Zaremba, W.; Szlam, A.; and LeCun, Y.
\newblock 2013.
\newblock Spectral networks and locally connected networks on graphs.
\newblock {\em arXiv:1312.6203}.

\bibitem[\protect\citeauthoryear{Cao \bgroup et al\mbox.\egroup
  }{2015}]{Cao2015}
Cao, C.; Bradley, D.; Zhou, K.; and Beeler, T.
\newblock 2015.
\newblock Real-time high-fidelity facial performance capture.
\newblock {\em ACM Trans. Graph.} 34(4):46.

\bibitem[\protect\citeauthoryear{Choy \bgroup et al\mbox.\egroup
  }{2016}]{choy20163d}
Choy, C.~B.; Xu, D.; Gwak, J.; Chen, K.; and Savarese, S.
\newblock 2016.
\newblock {3D-R2N2}: A unified approach for single and multi-view {3D} object
  reconstruction.
\newblock In {\em ECCV},  628--644.

\bibitem[\protect\citeauthoryear{Crane, Weischedel, and
  Wardetzky}{2013}]{Crane:2013:GH}
Crane, K.; Weischedel, C.; and Wardetzky, M.
\newblock 2013.
\newblock {Geodesics in Heat: A New Approach to Computing Distance Based on
  Heat Flow}.
\newblock {\em ACM Trans. Graph.} 32.

\bibitem[\protect\citeauthoryear{Defferrard, Bresson, and
  Vandergheynst}{2016}]{defferrard2016convolutional}
Defferrard, M.; Bresson, X.; and Vandergheynst, P.
\newblock 2016.
\newblock Convolutional neural networks on graphs with fast localized spectral
  filtering.
\newblock In {\em NIPS},  3844--3852.

\bibitem[\protect\citeauthoryear{Duvenaud \bgroup et al\mbox.\egroup
  }{2015}]{duvenaud2015convolutional}
Duvenaud, D.~K.; Maclaurin, D.; Iparraguirre, J.; Bombarell, R.; Hirzel, T.;
  Aspuru-Guzik, A.; and Adams, R.~P.
\newblock 2015.
\newblock Convolutional networks on graphs for learning molecular fingerprints.
\newblock In {\em NIPS},  2224--2232.

\bibitem[\protect\citeauthoryear{Gao \bgroup et al\mbox.\egroup
  }{2016}]{Gao2016}
Gao, L.; Lai, Y.-K.; Liang, D.; Chen, S.-Y.; and Xia, S.
\newblock 2016.
\newblock Efficient and flexible deformation representation for data-driven
  surface modeling.
\newblock {\em ACM Trans. Graph.} 35(5):158.

\bibitem[\protect\citeauthoryear{{Gao} \bgroup et al\mbox.\egroup
  }{2017}]{Gao2017}
{Gao}, L.; {Lai}, Y.-K.; {Yang}, J.; {Zhang}, L.-X.; {Kobbelt}, L.; and {Xia},
  S.
\newblock 2017.
\newblock {Sparse Data Driven Mesh Deformation}.
\newblock {\em arXiv:1709.01250}.

\bibitem[\protect\citeauthoryear{Gao, Zhang, and Lai}{2012}]{Gao2012}
Gao, L.; Zhang, G.; and Lai, Y.
\newblock 2012.
\newblock Lp shape deformation.
\newblock {\em Science China Information Sciences} 55(5):983--993.

\bibitem[\protect\citeauthoryear{Girdhar \bgroup et al\mbox.\egroup
  }{2016}]{Girdhar16b}
Girdhar, R.; Fouhey, D.; Rodriguez, M.; and Gupta, A.
\newblock 2016.
\newblock Learning a predictable and generative vector representation for
  objects.
\newblock In {\em ECCV}.

\bibitem[\protect\citeauthoryear{Guo, Zou, and Chen}{2015}]{Guo2015}
Guo, K.; Zou, D.; and Chen, X.
\newblock 2015.
\newblock {3D} mesh labeling via deep convolutional neural networks.
\newblock {\em ACM Trans. Graph.} 35(1):3.

\bibitem[\protect\citeauthoryear{Havaldar}{2006}]{Havaldar2006}
Havaldar, P.
\newblock 2006.
\newblock Performance driven facial animation.
\newblock In {\em ACM SIGGRAPH 2006 Course 30 Notes}.

\bibitem[\protect\citeauthoryear{Huang \bgroup et al\mbox.\egroup
  }{2014}]{huang}
Huang, Z.; Yao, J.; Zhong, Z.; Liu, Y.; and Guo, X.
\newblock 2014.
\newblock Sparse localized decomposition of deformation gradients.
\newblock {\em Comp. Graph. Forum} 33(7):239--248.

\bibitem[\protect\citeauthoryear{Kavan, Sloan, and
  O'Sullivan}{2010}]{CGF:CGF1602}
Kavan, L.; Sloan, P.-P.; and O'Sullivan, C.
\newblock 2010.
\newblock Fast and efficient skinning of animated meshes.
\newblock {\em Comp. Graph. Forum} 29(2):327--336.

\bibitem[\protect\citeauthoryear{Kingma and Ba}{2015}]{Kingma2015}
Kingma, D., and Ba, J.
\newblock 2015.
\newblock {ADAM}: A method for stochastic optimization.
\newblock In {\em ICLR}.

\bibitem[\protect\citeauthoryear{Li \bgroup et al\mbox.\egroup
  }{2015}]{li2015joint}
Li, Y.; Su, H.; Qi, C.~R.; Fish, N.; Cohen-Or, D.; and Guibas, L.~J.
\newblock 2015.
\newblock Joint embeddings of shapes and images via cnn image purification.
\newblock {\em ACM Trans. Graph.} 34(6):234.

\bibitem[\protect\citeauthoryear{Li \bgroup et al\mbox.\egroup
  }{2017}]{li_sig17}
Li, J.; Xu, K.; Chaudhuri, S.; Yumer, E.; Zhang, H.; and Guibas, L.
\newblock 2017.
\newblock Grass: Generative recursive autoencoders for shape structures.
\newblock {\em ACM Trans. Graph.} 36(4).

\bibitem[\protect\citeauthoryear{Maturana and
  Scherer}{2015}]{maturana2015voxnet}
Maturana, D., and Scherer, S.
\newblock 2015.
\newblock Voxnet: a {3D} convolutional neural network for real-time object
  recognition.
\newblock In {\em IEEE Conference on Intelligent Robots and Systems},
  922--928.

\bibitem[\protect\citeauthoryear{Nash and Williams}{2017}]{nash2017shape}
Nash, C., and Williams, C.~K.
\newblock 2017.
\newblock The shape variational autoencoder: A deep generative model of
  part-segmented {3D} objects.
\newblock {\em Comp. Graph. Forum}.

\bibitem[\protect\citeauthoryear{Neumann \bgroup et al\mbox.\egroup
  }{2013}]{neumann2013sparse}
Neumann, T.; Varanasi, K.; Wenger, S.; Wacker, M.; Magnor, M.; and Theobalt, C.
\newblock 2013.
\newblock Sparse localized deformation components.
\newblock {\em ACM Trans. Graph.} 32(6):179.

\bibitem[\protect\citeauthoryear{Niepert, Ahmed, and
  Kutzkov}{2016}]{niepert2016learning}
Niepert, M.; Ahmed, M.; and Kutzkov, K.
\newblock 2016.
\newblock Learning convolutional neural networks for graphs.
\newblock In {\em ICML},  2014--2023.

\bibitem[\protect\citeauthoryear{Sharma, Grau, and
  Fritz}{2016}]{sharma2016vconv}
Sharma, A.; Grau, O.; and Fritz, M.
\newblock 2016.
\newblock Vconv-dae: Deep volumetric shape learning without object labels.
\newblock In {\em ECCV Workshops},  236--250.

\bibitem[\protect\citeauthoryear{Shi \bgroup et al\mbox.\egroup
  }{2015}]{shi2015deeppano}
Shi, B.; Bai, S.; Zhou, Z.; and Bai, X.
\newblock 2015.
\newblock Deeppano: Deep panoramic representation for 3-d shape recognition.
\newblock {\em IEEE Signal Processing Letters} 22(12):2339--2343.

\bibitem[\protect\citeauthoryear{Sinha \bgroup et al\mbox.\egroup
  }{2017}]{sinha2017surfnet}
Sinha, A.; Unmesh, A.; Huang, Q.; and Ramani, K.
\newblock 2017.
\newblock {SurfNet}: Generating {3D} shape surfaces using deep residual
  networks.
\newblock In {\em CVPR}.

\bibitem[\protect\citeauthoryear{Su \bgroup et al\mbox.\egroup
  }{2015}]{su2015multi}
Su, H.; Maji, S.; Kalogerakis, E.; and Learned-Miller, E.
\newblock 2015.
\newblock Multi-view convolutional neural networks for {3D} shape recognition.
\newblock In {\em IEEE ICCV},  945--953.

\bibitem[\protect\citeauthoryear{Sumner and
  Popovi\'{c}}{2004}]{Sumner:2004:DTT:1015706.1015736}
Sumner, R.~W., and Popovi\'{c}, J.
\newblock 2004.
\newblock Deformation transfer for triangle meshes.
\newblock {\em ACM Trans. Graph.} 23(3):399--405.

\bibitem[\protect\citeauthoryear{Sumner \bgroup et al\mbox.\egroup
  }{2005}]{Sumner2005}
Sumner, R.~W.; Zwicker, M.; Gotsman, C.; and Popovi\'{c}, J.
\newblock 2005.
\newblock Mesh-based inverse kinematics.
\newblock {\em ACM Trans. Graph.} 24(3):488--495.

\bibitem[\protect\citeauthoryear{Tena, De~la Torre, and
  Matthews}{2011}]{Tena2011}
Tena, J.~R.; De~la Torre, F.; and Matthews, I.
\newblock 2011.
\newblock Interactive region-based linear {3D} face models.
\newblock {\em ACM Trans. Graph.} 30(4):76.

\bibitem[\protect\citeauthoryear{Tulsiani \bgroup et al\mbox.\egroup
  }{2016}]{tulsiani2016learning}
Tulsiani, S.; Su, H.; Guibas, L.~J.; Efros, A.~A.; and Malik, J.
\newblock 2016.
\newblock Learning shape abstractions by assembling volumetric primitives.
\newblock {\em arXiv:1612.00404}.

\bibitem[\protect\citeauthoryear{Vasa and Skala}{2011}]{5416707}
Vasa, L., and Skala, V.
\newblock 2011.
\newblock A perception correlated comparison method for dynamic meshes.
\newblock {\em IEEE Trans. Vis. Comp. Graph.} 17(2):220--230.

\bibitem[\protect\citeauthoryear{Vlasic \bgroup et al\mbox.\egroup
  }{2008}]{Vlasic2008}
Vlasic, D.; Baran, I.; Matusik, W.; and Popovi\'{c}, J.
\newblock 2008.
\newblock Articulated mesh animation from multi-view silhouettes.
\newblock {\em ACM Trans. Graph.} 27(3):97.

\bibitem[\protect\citeauthoryear{Wang \bgroup et al\mbox.\egroup }{2016}]{wang}
Wang, Y.; Li, G.; Zeng, Z.; and He, H.
\newblock 2016.
\newblock Articulated-motion-aware sparse localized decomposition.
\newblock {\em Comp. Graph. Forum}.

\bibitem[\protect\citeauthoryear{Wu \bgroup et al\mbox.\egroup
  }{2015}]{wu20153d}
Wu, Z.; Song, S.; Khosla, A.; Yu, F.; Zhang, L.; Tang, X.; and Xiao, J.
\newblock 2015.
\newblock {3D} {ShapeNets}: A deep representation for volumetric shapes.
\newblock In {\em CVPR},  1912--1920.

\bibitem[\protect\citeauthoryear{Wu \bgroup et al\mbox.\egroup }{2016}]{3dgan}
Wu, J.; Zhang, C.; Xue, T.; Freeman, W.~T.; and Tenenbaum, J.~B.
\newblock 2016.
\newblock Learning a probabilistic latent space of object shapes via {3D}
  generative-adversarial modeling.
\newblock In {\em NIPS},  82--90.

\bibitem[\protect\citeauthoryear{Yan \bgroup et al\mbox.\egroup
  }{2016}]{NIPS2016_6206}
Yan, X.; Yang, J.; Yumer, E.; Guo, Y.; and Lee, H.
\newblock 2016.
\newblock Perspective transformer nets: Learning single-view 3d object
  reconstruction without 3d supervision.
\newblock In {\em NIPS}.
\newblock  1696--1704.

\bibitem[\protect\citeauthoryear{Yi \bgroup et al\mbox.\egroup
  }{2017}]{Yi_2017_CVPR}
Yi, L.; Su, H.; Guo, X.; and Guibas, L.~J.
\newblock 2017.
\newblock {SyncSpecCNN}: Synchronized spectral {CNN} for {3D} shape
  segmentation.
\newblock In {\em CVPR}.

\bibitem[\protect\citeauthoryear{Zhang \bgroup et al\mbox.\egroup
  }{2004}]{zhang-siggraph2004-stfaces}
Zhang, L.; Snavely, N.; Curless, B.; and Seitz, S.~M.
\newblock 2004.
\newblock Spacetime faces: High-resolution capture for modeling and animation.
\newblock In {\em ACM SIGGRAPH},  548--558.

\bibitem[\protect\citeauthoryear{Zou, Hastie, and Tibshirani}{2004}]{zou2004}
Zou, H.; Hastie, T.; and Tibshirani, R.
\newblock 2004.
\newblock Sparse principal component analysis.
\newblock {\em J. Comp. Graph. Statistics} 15:2006.

\end{thebibliography}
\bibliographystyle{aaai}
\end{document}